\documentstyle[aps,twocolumn,tighten,epsfig]{revtex}
\begin{document}

\draft
\title{ DENSITY OF KINKS JUST AFTER A QUENCH IN AN UNDERDAMPED SYSTEM }

\author{Jacek Dziarmaga\thanks{E-mail: {\tt ufjacekd@thrisc.if.uj.edu.pl}}}
\address{ Institute of Physics, Jagiellonian University,
          Reymonta 4, 30-059 Krak\'ow, Poland}
\date{April 26, 1998}
\maketitle
\tighten

\begin{abstract}
{\bf A quench in an underdamped one dimensional $\phi^4$ model is studied
by analytical methods. The density of kinks just after the
transition is proportional to the square root of the rate of the quench
for slow quenches. If the quench is shorter that the relaxation time,
then the density scales like the third root of the rate. }
\end{abstract}
\vspace*{0.5cm}

   Topological defects play a prominent role in many condensed matter systems,
see e.g.\cite{davydov} for a review. It was also suggested that they
were an important ingredient of the early universe \cite{vilenkin}.
Topological defects can be generated in large numbers during
a second order phase transition. The dynamics of such a transition
to a symmetry broken phase has been an object of much recent attention
because of its importance in the cosmological context \cite{cosmos} and in
condensed matter physics \cite{condmat}. An early estimate of defect density
after a quench was given by Kibble \cite{kibble}. It his theory, the speed
of light is a dominant factor which determines the size of correlated domains.
Another scenario was put forward by Zurek \cite{zurek}, who emphasized the
importance of the nonequilibrium dynamics of the order parameter. Recent
experiments point to the latter theory.

   The key observation we make in this communication is that the relaxation
time in the underdamped limit is the same for all Fourier modes and it is
an inverse of the viscosity coefficient $\gamma$. If a quench is slow enough
so that the system can equilibrate during the initial stage of the quench,
then this relaxation time determines the frozen correlation length
and thus the density of kinks just after a quench. If a quench is
faster than the relaxation time, then there is not enough time to
equilibrate and the system developes a characteristic length scale
out of equlibrium.

   To be more specific, let us consider a one dimensional underdamped
$\phi^4$ model

\begin{eqnarray}\label{model}
&& \ddot{\phi}(t,x)+\gamma\dot{\phi}(t,x) = \nonumber \\
&& \phi''(t,x) - 2 a(t) \phi(t,x) - 2 \phi^{3}(t,x) + \eta(t,x)    \;\;,
\end{eqnarray}
where $\dot{}\equiv\partial{t}$, $ '\equiv\partial_{x}$. $\phi(t,x)$ is
a real order parameter. The system is assumed to be strongly underdamped
for the relevant part of the quench, $2a(t)>>\frac{\gamma^2}{4}$.
$\eta(t,x)$ is a Gaussian noise of temperature $T$ with correlations

\begin{eqnarray}\label{correlations}
&& <\eta(t,x)>=0 \;\;, \nonumber\\
&& <\eta(t_1,x_1)\eta(t_2,x_2)>=
   2 \gamma T \delta(t_1-t_2)\delta(x_1-x_2) \;\;.
\end{eqnarray}
The coefficient $a(t)$ is time dependent. We consider a linear quench

\begin{equation}\label{a(t)}
a(t)=
\left\{
\begin{array}{ll}
A                   & \mbox{ , if $\; t<0$ }                   \\
A(1-\frac{t}{\tau}) & \mbox{ , if $\; 0<t<\tau\frac{A+1}{A}$ } \\
-1                  & \mbox{ , if $\; \tau\frac{A+1}{A}<t$ }
\end{array}
\right. 
\end{equation}
Before the quench, for $t<0$, the system is in a symmetric phase $(A>0)$,
during the quench, at $t=\tau$, it undergoes a transition from the symmetric
phase $a(t<\tau)>0$ to a broken symmetry phase $a(t>\tau)<0$. Finally it
settles down at $a(t)=-1$.

  As long as the system is still in the symmetric phase ($t<\tau$)
and the temperature is moderate, the field $\phi$ can be regarded as a
small fluctuation around its symmetric ground state $<\phi(t,x)>=0$. It is
justified to neglect the cubic term on the RHS of Eq.(\ref{model}).
The field can be written as a Fourier transform,

\begin{equation}\label{fourier}
\phi(t,x)=\int_{-\infty}^{+\infty}dk\; \tilde{\phi}(t,k) e^{ikx} \;\;.
\end{equation}
The Fourier transform of linearized Eq.(\ref{model}) and of the noises
(\ref{correlations}) is

\begin{eqnarray}\label{model1}
&& \ddot{\tilde{\phi}}(t,k)+\gamma\dot{\tilde{\phi}}(t,k)=
   -[2 a(t)+k^2]\tilde{\phi}(t,k)+\tilde{\eta}(t,k) \;\;, \nonumber \\
&& <\tilde{\eta}(t,k)>=0  \;\;,\nonumber \\
&& <\tilde{\eta}^{*}(t_1,k_1)\tilde{\eta}(t_2,k_2)>=
   \frac{\gamma T}{\pi}\delta(t_1-t_2)\delta(k_1-k_2) \;\;.
\end{eqnarray}
To predict the density of kinks in the symmetric phase we must know the two
point correlation function
$<\tilde{\phi}^{*}(\tau,k_1)\tilde{\phi}(\tau,k_2)>$ at the critical point
$t=\tau$. An {\it exact} formal expression can be constructed for any
$\gamma$ as a rather involved combination of Bessel functions but it is
too complicated to be very illuminating. We concentrate here on the limit
of very small $\gamma$, where some intuitive understanding can be achieved.

  The limit $\gamma\rightarrow 0$ is a delicate matter as $\gamma$ appears
both in the field equations and in the noise correlations. Any {\it ad hoc}
setting of $\gamma=0$ here or there may be dangerous. Our approach
to this problem is as follows. In the strongly underdamped limit
the system (\ref{model1}) can be mapped to the equivalent one

\begin{eqnarray}\label{model2}
&& \dot{\tilde{\phi}}(t,k)=
   -\frac{\gamma}{2}\tilde{\phi}(t,k)+n(t,k) \;\;, \nonumber \\
&& <n(t,k)>=0  \;\;,\nonumber \\
&& <n(t_1,k_1) n(t_2,k_2)>=
   \frac{\gamma T \delta(t_1-t_2) \delta(k_1-k_2) }{2\pi [2a(t_1)+k^2_1]}\;\;.
\end{eqnarray}
If $a(t)=a=const$, then both systems have correlations

\begin{equation}
<\tilde{\phi}^{*}(t,k_1)\tilde{\phi}(t,k_2)>=
      \frac{T\delta(k_1-k_2)}{2\pi [2a+k^2_1]}      \;\;.
\end{equation}
Their response to the change of parameters is the same too. In particular
they have the same relaxation time. For example, if $a(t)=a=const$ and the
temperature jumps from $T=0$ for $t<0$ to $T>0$ for $t>0$, then the
correlation functions for $t>0$ are

\begin{eqnarray}
&&<\tilde{\phi}^{*}(t,k_1)\tilde{\phi}(t,k_2)>=
  \frac{T\delta(k_1-k_2)}{2\pi [2a+k^2_1]} [1-e^{-\gamma t}] \;\;,\nonumber\\
&&<\tilde{\phi}^{*}(t,k_1)\tilde{\phi}(t,k_2)>= \nonumber\\
&&\;\;\frac{T\delta(k_1-k_2)}{2\pi [2a+k^2_1]}
      [1-e^{-\gamma t}+e^{-\gamma t}O(\frac{\gamma}{\sqrt{2a+k^2}})]
\end{eqnarray}
for the models (\ref{model2}) and (\ref{model1}) respectively.
$O(\frac{\gamma}{\sqrt{2a+k^2}})$ denotes oscillating terms which
are negligible for small $\gamma$. In this limit the response
of the two systems is the same. Another example is the response
to a sudden change in the parameter $a$ from $a=a_{-}$ for $t<0$
to $a=a_{+}$ for $t>0$

\begin{eqnarray}
&&<\tilde{\phi}^{*}(t,k_1)\tilde{\phi}(t,k_2)>=
  \frac{T\delta(k_1-k_2)}{2\pi}
  [\frac{1-e^{-\gamma t}}{2a_{+}+k^2}+
   \frac{e^{-\gamma t}}{2a_{-}+k^2}   ] \;\;,\nonumber\\
&&<\tilde{\phi}^{*}(t,k_1)\tilde{\phi}(t,k_2)>=
  \frac{T\delta(k_1-k_2)}{2\pi}\times  \nonumber\\
&&\;\; [\frac{1-e^{-\gamma t}}{2a_{+}+k^2}+
        \frac{e^{-\gamma t}}{2a_{-}+k^2}+
        O(\frac{\gamma e^{-\gamma\tau}}{\sqrt{2a_{+}+k^2}},
          \frac{\gamma e^{-\gamma\tau}}{\sqrt{2a_{-}+k^2}})   ] \;\;,
\end{eqnarray}
where the $O(\ldots)$ terms are once again oscillations negligible
for $\gamma\rightarrow 0$. The systems respond in
the same way to an impulsive perturbation of their parameters. The
impulsive perturbation contains all frequencies so the models
are equivalent in the underdamped limit.

   From now on we work in the framework of the model (\ref{model2}).
The formal solution of Eqs.(\ref{model2}) is
\begin{equation}
\tilde{\phi}(\tau,k)=\int_{-\infty}^{\tau}dt_1\;
                     e^{-\gamma (\tau-t_1)/2} n(t_1,k) \;\;.
\end{equation}
This solution, the correlations (\ref{model2}) and the explicit form of
$a(t)$, see Eq.(\ref{a(t)}), give the correlation function at $t=\tau$

\begin{eqnarray}\label{phiphi}
&& <\tilde{\phi}^{*}(\tau,k_1)\tilde{\phi}(\tau,k_2)>=
   \frac{T}{2\pi}\delta(k_1-k_2)\;G(\tau,k_1)         \nonumber\\
&& G(\tau,k)=\frac{e^{-\gamma\tau}}{2A+k^2}+            
   \int_{0}^{\gamma\tau}dx\; \frac{e^{-x}}{\frac{2A}{\gamma\tau}x+k^2}\;\;.
\end{eqnarray}
$G(\tau,k)$ has a very interesting asymptote for $\tau>>1/\gamma$

\begin{equation}\label{Gasymptote}
G(\tau,k)\approx\frac{1}{\frac{2A}{\gamma\tau}+k^2} \;\;.
\end{equation}
This asymptote can be understood in heuristic way. In the extreme
underdamped limit (\ref{model2}) all the Fourier modes have
the same relaxation time $1/\gamma$ whatever is their momentum $k$.
After the quench begins at $t=0$ the system tries to adapt to the changes
of the parameter $a(t)$. It can not react on a time scale shorter
than $1/\gamma$. The correlations of the system are effectively
"frozen" at the instant $\hat{t}$, when the time still left to the
transition at $t=\tau$ is equal to the relaxation time,
$(\tau-\hat{t})=1/\gamma$. At this instant $a(\hat{t})=\frac{A}{\gamma\tau}$
and the correlation function is frozen as $(\ref{Gasymptote})$.
The frozen correlation length is $\sqrt{\frac{\gamma\tau}{2A}}$
so the density of kinks just after a quench scales like

\begin{equation}
n \sim \sqrt{\frac{A}{\gamma\tau}}
\end{equation}
provided that the quench time is longer than the relaxation time

\begin{equation}
\tau >> \frac{1}{\gamma}
\end{equation}
and it is short enough so that the system is still underdamped
at the freeze-in time $\hat{t}$, $2a(\hat{t}) >> \gamma^2/4$ or

\begin{equation}
\tau << \frac{8A}{\gamma^3}
\end{equation}
The range of quench times $\tau$ where the $\tau^{-1/2}$ scaling applies
blows up as $\gamma$ tends to zero.

   If the quench time is shorter than the relaxation time $\tau<<1/\gamma$
then the system is out of equlibrium from the very beginning of the
quench. As the system does not equilibrate anyway, we can effectively
set $\gamma=0$ in Eqs.(\ref{model},\ref{model1}). In the absence
of damping exact solutions of equations (\ref{model1}) for the linear
quench (\ref{a(t)}) are combinations of the modes

\begin{eqnarray}\label{freeroll}
&& \tilde{\phi}_{k}(0<t<\tau) \sim                  \nonumber\\
&& \sqrt{ \tau (1+\frac{k^2}{2A}) - t } \;\;
   J_{\stackrel{+}{-}\frac{1}{3}}
   [\frac{2}{3}\sqrt{\frac{2A}{\tau}}\;\;
   (\tau (1+\frac{k^2}{2A}) - t)^{\frac{3}{2}}]\;\;.
\end{eqnarray}
At the critical point $t=\tau$ the modes take the form

\begin{equation}\label{freerolltau}
\tilde{\phi}_{k}(t=\tau) \sim
\sqrt{ \frac{\tau k^2}{2A} } \;\;
J_{\stackrel{+}{-}\frac{1}{3}}
[\frac{\tau k^3}{3A}]\;\;.
\end{equation}
The correlation function at $t=\tau$ can be constructed as a combination
of $\tilde{\phi}_{k}(0),\dot{\tilde{\phi}}_{k}(0)$ and of
$\tilde{\phi}_{k}(\tau)$. The latter contain information about a characteristic 
length scale at the critical point. At $\frac{\tau k^3}{3A}\approx 1$ the 
behaviour of the Bessel function (\ref{freerolltau}) changes qualitatively. An 
inverse of this length scale determines the density of kinks just after a 
quench

\begin{equation}
n \sim (\frac{A}{\tau})^\frac{1}{3}
\end{equation}
provided that

\begin{equation}
\tau<<\frac{1}{\gamma} \;\;.
\end{equation}
This result is consistent with the discussion of Laguna and Zurek 
\cite{zu} and with results of Karra and Rivers \cite{kr}. The opposite 
$\tau>>1/\gamma$ regime has not been discussed before.

\subsection*{Conclusion.}

For fast underdamped quenches the frozen correlation length grows like 
$\tau^{1/3}$. For slow underdamped quenches it scales like $\tau^{1/2}$.

\acknowledgements

Discussions with Wojciech Zurek are much appreciated.

\end{document}